\documentclass{sig-alternate-05-2015}
\usepackage{listings, color}
\usepackage{cite}
\definecolor{forestgreen}{RGB}{34,139,34}
\definecolor{orangered}{RGB}{239,134,64}
\definecolor{darkblue}{rgb}{0.0,0.0,0.6}
\definecolor{gray}{rgb}{0.4,0.4,0.4}
\definecolor{backcolour}{rgb}{0.95,0.95,0.95}
\lstdefinestyle{XML} {
    language=XML,
    extendedchars=true, 
    breaklines=true,
    breakatwhitespace=true,
    emph={},
    emphstyle=\color{red},
    basicstyle=\ttfamily,
    columns=fullflexible,
    commentstyle=\color{gray}\upshape,
    morestring=[b]",
    morecomment=[s]{<?}{?>},
    morecomment=[s][\color{forestgreen}]{<!--}{-->},
    keywordstyle=\color{orangered},
    stringstyle=\ttfamily\color{black}\normalfont,
    tagstyle=\color{darkblue}\bf,
    backgroundcolor=\color{backcolour},
    morekeywords={name,aref,id,iref,content,type},
    keepspaces=true,                 
    numbersep=5pt,                  
    showspaces=false,                
    showstringspaces=false,
    showtabs=false,                  
    tabsize=1
}
\makeatletter
\def\@copyrightspace{\relax}
\makeatother
\begin{document}

\setcopyright{acmcopyright}

\title{Model based approach to study Defect Dependency in Large Scale Integrated Software Products}

\numberofauthors{2} 
\author{
%
%
\alignauthor
Sai Anirudh Karre\\
       \affaddr{Software Engineering Research Center}\\
       \affaddr{IIIT Hyderabad, India}\\
        \affaddr{sai.anirudh@research.iiit.ac.in}
\alignauthor
Y. Raghu Reddy\\
       \affaddr{Software Engineering Research Center}\\
       \affaddr{IIIT Hyderabad, India}\\
        \affaddr{raghu.reddy@iiit.ac.in}
}

\maketitle
\begin{abstract}
Large organizations have diverse product offerings to meet various business needs. To increase revenue, its common these days to offer software products as integrated product suite(s) rather than individual products. Creating and maintaining high quality software products within the integrated product suite requires rigorous product engineering methods. The sheer size of products and dependencies involved tend to raise unidentified defects that may become critical post product upgrades or after every release cycle. It is difficult to track such defects and its widespread across underlying sub-products. In this paper, we present a model based approach to study the defect dependency in large scale integrated software products to avoid surprise defects after product release. To validate the approach, we have applied it on some pilot projects in industry.
\end{abstract}

\keywords{Defect Dependency, Software Quality Analysis, Integrated Software Products, XML, UML, \textit{i*} modeling framework}

\section{Introduction}
Quality of software is a major factor in determining the success of software products. In November 2014, Aplhr.com (a leading technology website) published an article\footnote{http://bit.ly/1WAMSiB} headlined: \textit{``Microsoft fixes 19-year-old Windows bug - but what about XP?"} describes the release of patches to the critical bugs that have been laying dormant in one of the world's leading operating system - Windows\texttrademark for about two decades. In January 2015, Softpedia.com (a technology website) published an article\footnote{http://bit.ly/1UqtlRz} headlined: \textit{``Why iOS and OS X Today Are Buggy?"} and znet.com (a technology website) published another article\footnote{http://zd.net/1raiUWs} headlined: \textit{``Apple has a serious problem with software quality"} describes deep insights on challenges faced by major technology giants like Apple on failing to deliver quality products to its customers. Such instances substantiate the importance of quality in large scale software products, especially those that consist of multiple releases and consist of multiple sub-products. Quality of a software is ideally expected to improve over multiple product/sub-product releases with latest release being the most stable one.

In the recent years, large organizations have offered multiple products/sub-products as integrated product suites to cater to the business needs of their clients. The sub-products are interconnected to each other and share common packages of program to form one application. Integrated product suites tend to be more complex in design than non-integrated software products due to the various inter connections and dependencies between the products/sub-products. As a result, integrated product suites require detailed exploration of the widespread of a defect across the entire product suite along with causal analysis so as to improve overall product quality.   However, at times, either due to security constraints or due to complicated work-flow, quality engineers may not be able to access and analyze the source code of all sub-products together in the product suite. Thus investigating \textit{defect dependency} (an indicator to identify defect widespread) in large scale integrated software product is difficult \cite{Sai}.

In large scale integrated software products, a small proportion of defects are fixed as soon as they are reported. Rest of the defects are targeted for fixes in future product release cycles. 
Among such targeted defects, a high number of them are insignificant and innocuous in the current version but may have the potential to become acute in future versions. As per Gartner's 2015 Magic Quadrant for Enterprise Integration platforms as a Service Survey \cite{Gartner}, most of Software manufactures that develop complex integrated products still depend on traditional approaches to maintain quality standards of their existing products. The report states that new trends in research are tougher to adopt in current integrated product ecosystem mainly due to constraints like critical release deadlines, resource availability, product scalability, and their aversion towards risk of implementing research results. The research presented in this paper is motivated by the following facts: (1) A stable software product can evolve over various version releases if and only if efficient quality measures are adapted, and (2) Simple approaches are required to estimate defect dependency in Large Scale Integrated software products to address imminent software quality issues.

In industry, code inspections and dynamic analysis are standard ways to analyse defect dependencies arising from data flow and control flow in a given source code. With increase in number of lines of code, code inspection becomes burdensome to study the defect flow. In case of large scale integrated software products, source code analysis is even more difficult as most of the sub-products are integrated after being developed on different platforms. In such situations, model analysis of a sub-product fills the gap on understanding data flow and control flow across the products/sub-products. 

Enterprise software tools like SciTools, McCabe and IBM ClearQuest provide Code Dependency Analysis in connected systems with limited programming language support. However they do not provide defect dependency analysis on disconnected integrated software systems. This triggered us to analyse and utilize models as the primary artifact to study defect dependency. There are various modeling frameworks and underlying languages that provide the flexibility to record relationship between elements defined as per software product design. \textit{i*} modeling framework is one such advanced modeling framework which efficiently captures relationships between elements \cite{Eric}. 

In this paper, we propose a simple model based approach using \textit{i*} modeling framework to calculate defect dependency in a large scale integrated software products so as to improve software quality and help Product Manager to prioritize defects based on degree of the metric. We have applied the approach over a pilot projects to validate and verify the approach. We summarize the real-time requirements and challenges associated with the approach. 

The rest of the paper is organized as follows: Section II provides background to our work, section III discusses some related work. Section IV describes the proposed model based approach and steps to calculate the defect dependency. Section V describes the implementation of our approach on industrial pilot study, along with the key takeaways and challenges. Section VI discusses threats to validity. Finally,  section VII provides conclusion and some insights about the future work.

\section{Background}

\subsection{Why Large Scale Integrated Software Products?}
Small organizations, start-ups or new development projects have the potential to implement new trends/approaches based on research in Software Quality. It is a challenge for larger organizations with well-established products to adhere to these changes as it requires massive planning and human effort, especially in case of \textit{\textbf{Integrated Software Product}}. In large scaled integrated products, where the sub-products are referred to as \textit{`product pillars'}, the product becomes vulnerable if its product pillars are bounded with integration defects. For example, let us consider an integrated software product consisting of the following two sub-products: \textit{Supply-Chain sub-product} and \textit{Revenue Reporter sub-product}. Supply-chain sub-product generally tracks product billing while revenue reporter sub-product reports revenue. A common defect in the integrated product is \textit{rounding-off of the product price}. If an entity in supply chain sub-product records a value of 34.55599 and if it is rounded off to 35, it may not cause a big change in the entire revenue balance sheet. However, if all the entities in the revenue balance sheet are rounded off to an absolute value, there would be an abrupt shift in revenues. As a result, from an integrated product perspective, the revenue reports show incorrect data. If the results are taken separately, rounding-off defect can be insignificant in chain-supply sub-product, but critical for product billing sub-product.  In such scenarios, the defect may be logged in different ways based on the product developer's assessment. For example, the same defect may be considered as a severe defect in revenue reporter sub-product where as it may not even be logged as a defect in supply-chain sub-product. Hence measuring the widespread of such defects across the products can be critical to the defect fix cycle and the release cycle \cite{Sai}.

During product development sprint cycles, software quality teams spend substantial effort and time on validating the fix over an evolving product. They ensure that the fix does not cause a new defect in existing sub-products. Rigorous checks are performed before releasing a validated fix of recorded defect(s). Integration issues may occur due to incorrect control flow and data flow between the products, sub-products or sub-modules within an entire large scale integrated software product and hence they are difficult to test. The stability of a large-scale integrated software product is directly proportional to its design and implementation quality \cite{Yoshida}. In such scenario, it is essential for product owners to understand the dependency of the defect so as to mitigate possible surprise defects from other modules of the large scale integrated software product.

\begin{table}[ht]
\centering
\caption{Defect Report - Stock Data Manager}
\label{def1}
\begin{center}
\begin{tabular} {|c| p{5cm}|} \hline
\textbf{Defect \#01} &Unknown validation errors while storing Stock Data in new Stock Portfolio module\\ \hline
\textbf{Module (\textit{Product})} &Stock Portfolio (Stock Data Manager)\\ \hline
\textbf{Cause} &Incompatible datatype defined for Stock Trend Value in new \textit{StockPortfolio} class\\ \hline
\textbf{Fix} &Updated closet possible datatype to overcome the issue. Handling expectations in case of error data\\ \hline
\end{tabular}
\end{center}
\end{table}
\begin{table}[ht]
\centering
\caption{Defect Report - Stock Predictor System}
\label{def2}
\begin{center}
\begin{tabular} {|c| p{5cm}|} \hline
\textbf{Defect\#02}&Rounding off Stock value causes Prediction error\\ \hline
\textbf{Module (\textit{Product})} &Stock-BI (Stock Predictor Systems)\\ \hline
\textbf{Cause}&New logic for trend stock data is rounding off stock value to 4 decimals which is causing errors\\ \hline
\textbf{Fix}&Provided provision to choose the round-off value for BI Team to visualize data in different ways\\ \hline
\end{tabular}
\end{center}
\end{table}
\begin{table}[ht]
\centering
\caption{Defect Report - Stock Pay Gateway}
\label{def3}
\begin{center}
\begin{tabular} {|c| p{5cm}|} \hline
\textbf{Defect\#03}&No validation warning for "In-sufficient Funds"\\ \hline
\textbf{Module (\textit{Product})} &Credit Payment (Stock Pay Gateway)\\ \hline
\textbf{Cause}&Missing Exception Handling for Credit Payments
\\ \hline
\textbf{Fix}&Change in Design to handle payment exceptions\\ \hline
\end{tabular}
\end{center}
\end{table}

In similar context, let us assume another example of a Stock Exchange Integrated Software product which was reconstructed using different independent legacy products \cite{Sai}. \textit{Stock Data Manager}, \textit{Stock Predictor Systems} and \textit{Stock Pay Gateway} are few of the pillar products whose sample defect reports are listed below. \textbf{\textit{Defect \#01}} is the defect recorded in \textit{Stock Portfolio} module of pillar product \textit{Stock Data Manager}, \textit{\textbf{Defect \#02}} is recorded in \textit{Stock-BI} module of pillar product \textit{Stock Predictor Systems} and \textit{\textbf{Defect \#03}} is recorded in \textit{Credit Payment} module of pillar product \textit{Stock Pay Gateway}. The description, cause and fix of the defects is given in table \ref{def1}, \ref{def2}, and \ref{def3}. If the product modules are considered individually, the defects may or may not be a major cause of concern. However, when these modules are integrated to form a large scale system, the flow of data and control may depend on one another leading to a different set of issues. 

Let us consider the case of \textbf{\textit{Defect \#01}} where in an incorrect stock portfolio data caused confusion in Business Intelligence teams as the defective stock data will result in incorrect consequences in critical corporate reporting dashboards. Also if a defective stock has Credit Payments issues as reported in \textbf{\textit{Defect \#03}}, it can result in digital fraud as per stock regulatory compliance law. \textbf{\textit{Defect \#02}} is more sensitive as it may give raise to money laundering issues in real time business scenario. These defects appear to be equally critical but we may need a scale to rank these defects using real dependency values over each other. In reality, it is a difficult task for Product Manager(s) of such large critical software to assess the widespread of a defect without studying the impact of one defect over another. However, with an understanding about defect widespread, Product Manager(s) will have an opportunity to prioritize defects so as to stabilize the product.

\subsection{Defect Dependency}
Defect Dependency is defined as an \textit{`indicator to identify the widespread of a defect with unknown impact and unknown risk over a module(s) or component(s) or sub-product(s) of a large scale complex software product'}. This can be represented as metric of quality or an indicator of dependency of a defect in Large Scale Integrated Software Products.

In our previous work \cite{Sai}, we presented a heuristics based approach to study defect dependency using rough set theory  and have implemented the approach on a real time industry defect data-set. We captured interesting observations across various version release cycles to understand product improvement. Our work required thorough implementation of the proposed metric across different release cycles to understand its significance and to assess the overall defect widespread. It was a time consuming approach as the successful implementation of the approach and the corresponding metric could only be done by comparing the results with previous product versions. 

Large Scale integrated software products have critical integrated work-flows, hence it is difficult to interpret dependencies via code and provide a value to represent the dependencies. As part of our research, we worked towards proposing a model based approach to estimate widespread of defect which covers dependencies across all the sub-products and help stakeholders to prioritize defects so as to establish a stable product.

\subsection{Modeling Languages}
\begin{figure} [h] 
\centering
\includegraphics{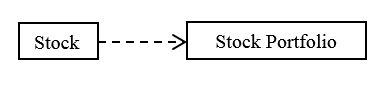} 
\caption{A simple class dependency relation in UML}
\label{depend}
\end{figure}

There are various modeling languages that provide flexibility to specify relationships i.e dependencies between entities defined as per design of the software product. A modeling language can be used to express information in a structure that is defined by a consistent set of rules. These rules are used for interpretation of the meaning of components in the structure. It is a simple way to represent requirements. There are many modeling languages, some of which like UML \cite{uml} are general-purpose modeling languages and some others are domain specific modeling languages. Kevoree Modeling framework, \textit{i*} modeling framework,  Service-oriented modeling framework (SOMF) etc. are few frameworks that use some underlying modeling language to facilitate modeling of particular applications/systems.  For example, if we consider two classes Stock and Stock Portfolio having a simple dependency relation, it can be represented using a Class diagram (shown in Fig:\ref{depend}), where the relationship indicates that a change to the Stock Portfolio class might require a change to the Stock class. The Connection between two packages is to indicate that at least one element in the Stock package is dependent on an element in the Stock Portfolio package. A modeling framework like Eclipse Modeling Framework (EMF) \cite{emf} can be used to create domain level applications, instances and perform some analysis on the instances of the model.

\section{Related Work}
Many researchers have proposed various approaches to address quality issues at various stages of software production. Clark et al. \cite{Clarke} were the first to study the program semantic and syntactic dependence based software testing for regular software development. It is one of the early studies on the widespread of defects at the code-level but the applicability of the approach on large code bases is questionable. Asgari et al. performed an empirical study on estimating software dependability \cite{Asgari} at architectural level. They worked mostly on improving integration architectures for large software. Unified Models of Dependability \cite{Victor} was proposed to identify dependencies at all levels of software development to extend defect dependency in terms of software quality. However, the primary goal was to stabilize the existing design and did not have a mechanism to extend the approach to existing software. Stormer et al. proposed required quality attributes in architecture reconstruction \cite{Stormer} which help developers to design a better architecture to avoid defect widespread. These proposed design strategies become unstable if they are applied to existing legacy software integration. There is a need for re-defining this work to fit large integrated software products. Nagappan et al. have come up with an industrial case study using software dependencies \cite{Ball} to study product failures. This work was limited to specific domain but opens up a wide opportunity to extend to large software.

Defect prediction is another way of studying defect wide-spread as it helps developers to understand what exactly can go wrong. Shihab \cite{Shihab} performed empirical analysis and found that there are more than 100 research papers published on Software Defect Prediction (SDP). However, most of the approaches do not provide guidance on industrial adoption or rarely consider the impact, risk and dependency associated with the predicted or forecasted defects. Practical adoption of SDP in industry to date is limited as software industry tends to be reactive than proactive. Most of the organizations need clear methodologies to identify most defective parts of their product and need recommendations to practitioners to prioritize defects so as to avoid breakdown of rest of the product.

Kitchnham et al. proposed a real-time model-driven approach called SQUID \cite{Kitchenham} to improve software quality in software production. Although the approach is considered to be pretty good, it is restricted to specific domain. Gotel et al. observed that large software products require generic methods and diversity in evaluating quality for industrial adoption \cite{Gotel}. They have proposed work-flow challenges involved in real time software integration projects. This helps research community to learn and understand the gravity of real-time industrial experiences and their concerns. Evaluation and measurement models for Quality were initially proposed by Tian \cite{Tian}. Tain's work covers standard processes required to evaluate the quality attributes using modeling. Wager et al. came up with a detailed study on integration approaches using quality modeling  and provided some insights on how integration challenges are to be addressed in incremental format \cite{Deissenboeck} \cite{Wagner}. Zain et al. \cite{Zain} have performed comparative analysis over all the existing software quality models and have captured their observations on how these models really contribute towards building better quality software. Overall, there are many model-based approaches suggested in literature to address quality issues but very few address the widespread of a defect or study dependency of a defect over an entire product. In addition, there are very limited practical implementations/case studies of model-based approaches that are oriented towards defect dependency.

\section{Proposed Approach}
\begin{figure} [h]
\centering
\includegraphics{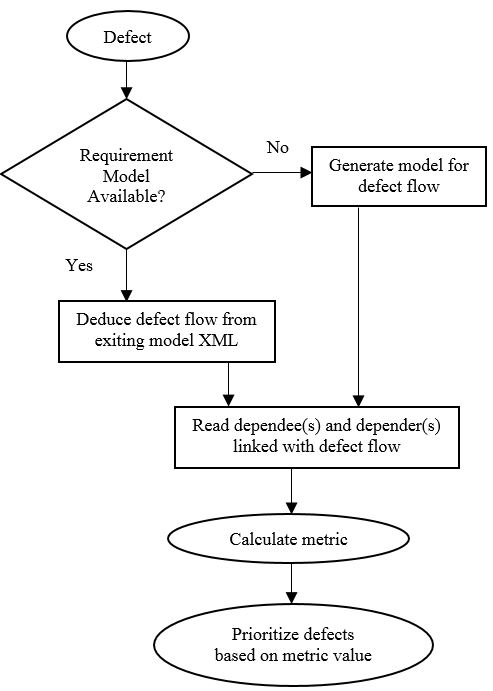}
\caption{Approach to calculate Defect Dependency}
\label{approach}
\end{figure}

This section contains details of our proposed model based approach to study the defect dependency of a defect across software module(s) or component(s) or sub-product(s) in large scale integrated software product.  Figure \ref{approach} provides an overview of the approach. We assume every defect is reported using some defect reporting form. Once a defect is identified/recorded by a stakeholder, its defect flow in existing product requirement model is reviewed. If there is no requirement model defined initially, at least a relative model which covers defect flow should be generated. Using the requirement model, the relationship between the entities are taken in consideration by fetching the details of depender, dependee and actors involved as part of defect flow. Using the formula specified in equation (\ref{eq:1}), the defect dependency metric value is determined. The product managers may then prioritize the defect fix cycle based on business need. The below subsections details the various aspects of the approach and also contains the detailed steps to calculate defect widespread.

\subsection{Depender, Dependee and Actors}
Consider \textit{A} and \textit{B} as two software elements (methods, class, object etc.) in a model holding a dependency relationship. If \textit{A} depends on \textit{B}, then \textit{A} is the \textit{\textbf{depender}} and \textit{B} is the \textit{\textbf{dependee}}. This notion exists in almost all standard modeling frameworks. Actors are the other elements that participate in the dependency relationship between a depender and dependee.

\subsection{Defect Flow}
Defects are reported and represented in the form of a Defect Report. If the products are from different organizations, they may use different defect reporting systems and the corresponding templates used for reporting the defects might also be different. However, almost all the defect tracking tools contain few standard attributes associated with a defect. Most prominent attributes are defect description, steps to reproduce the defect, module/product/sub-product and its severity. For an established large scale integrated software product, it is expected that a defect report is tracked using a Defect tracking system across various versions of a product. While developers provide fixes to the defects, quality engineers will have to validate the fix by understanding the defect flow across the defective module. They will have to revisit the use cases and build test cases based on known scenarios. It is a standard practice for a quality engineer to review the flow of defect across the product to assess the cause-effect issues. However, this exercise becomes tougher if multiple product/sub-product dependencies exist. As part of the approach, the first step is to capture the defect flow by mapping the defect attributes with the existing requirement model of an entire integrated software product. This model is used to identify the list of \textit{dependee(s)} and \textit{depender(s)} in regards to the reported defect along with the associated actors involved in dependency relation. This can be realized in two possible cases:

\begin{itemize}
\item\textit{Case A: } In case of new integrated software development projects, practitioners have to rely on the existing requirement model defined for the entire product suite so as to identify and record the dependee and depender associated with the reported defect. 

\item\textit{Case B: } In case of an existing integrated software projects with no proper requirement model across the sub-products, a requirement model needs to be constructed in regards to the defect flow across the sub-products. This requires additional effort from Product Management team as they have to understand the scope of data flow and control flow across the product. In this case, the requirement model will evolve over a period of time.
\end{itemize}

\subsection{Understanding XML File}
Almost all the modeling frameworks support XML standard to represent the requirement model. With the availability of a model XML file, \textit{dependee} and \textit{depender} can be identified based on the framework tag definition. The XML file instance of the entire integrated product has to be reviewed and the \textit{dependee(s)} and \textit{depender(s)} associated with the defect flow must be extracted. XML file instances can be easily generated by various modeling framework tools for the applications making use of the framework. One has to externally read the XML file instance and compute the count of desired elements to calculate the defect dependency value. The tags defined in the model XML file are not universal and vary from one modeling framework to another. While the XML file instance is reviewed and analyzed for the defect flow, it must be ensured that the file follows proper semantics and has valid tags based on the underlying modeling framework.

\subsection{Calculating Defect Dependency}
Using the model XML file of recorded defect and the model XML file of the entire product, we will apply below steps to calculate defect dependency:

\begin{itemize}
\item Consider \textit{d} to be a defect recorded in a large scale integrated product \textit{P}
\item Let \textit{c} be the notation for actor, \textit{e} be the notation for dependee and \textit{r} be the notation for depender in a model XML file.
\item Let \textit{dc} be the notation for count of actors involved in a defect \textit{d} and \textit{Pc} be the notation for count of actors in a entire Product \textit{P}
\item Read model XML file of \textit{d} and \textit{P} to count the actors \textit{(c)}, dependee (\textit{e}) and depender (\textit{r}) involved
\item Let \textit{dc} be the notation for sum of all actors involved in defect \textit{d} and \textit{Pc} be the notation for all the actors available in the entire product \textit{P}
\item Let \textit{de} be the notation for dependee in defect model and \textit{dr} be the notation for depender in defect model, then (\textit{de} + \textit{dr}) be sum of dependee and depender of a recorded defect \textit{d}
\item Let \textit{Pe} be the notation for dependee in entire product model and \textit{Pr} be the notation for depender in entire product model, then (\textit{Pe} + \textit{Pr}) be sum of dependee and depender of an entire product \textit{P}
\item Let $a = [dc * (de+dr)]$ be the spread of the entities dependent as per the defect i.e. the dependency weightage was evenly assigned to all the actors involved in the defect model. Similarly, let $b = [Pc * (Pe+Pr)]$ be the spread of entities dependent as per the overall product i.e. the dependency weightage was evenly assigned to all the actors involved in the overall product model.
\item Note that \textit{b} > \textit{a}, as the count of actors, dependee and depender are higher in the product model than in the defect model.
\item Hence (\textit{b} - \textit{a}) gives us the value of missing portion of \textit{a} in \textit{b} i.e. it represent the missing weightage of \textit{a} in \textit{b}. If we divide (\textit{b} - \textit{a}) by \textit{b}, we get the fraction weightage of missing portion of \textit{a} per unit weight of \textit{b}. This gives us the spread of missing portion of \textit{a} in \textit{b}
\item By subtracting the above value by 1, we gives us the weightage of a for per unit in \textit{b}. We represent the resultant defect dependency value as \textit{D}. Therefore the weightage of the defect widespread in
\textit{a} product is represented as below:
\begin{equation} \label{eq:1}
\textit{D}= 1 - (\frac{\textit{b}-\textit{a}}{\textit{b}})
\end{equation}
\item The scale of \textit{D} here is between (0,1) and it can also be represented in percentage by factoring it by 100
\item If $D=0$, the $\textit{b}-\textit{a}=\textit{b}$ i.e. the defect model do not contain any details of product model. If $D=1$, then $b-a=0$ i.e. the defect model is equal to the product model which is not common.
\end{itemize}

\subsection{Defect Prioritization}
Product Managers can utilize this metric value with a desired scale factor to prioritize the defect. The value \textit{D} is flexible to be customized as per functional needs, hence it can be round-off with any business factor. In day-to-day development, reconstruction and maintenance projects external properties like severity, customer criticality, impact factor and release deadline etc. are considered as additional pre-defined indicators to prioritize the defects. For example, most of the product based software industries classify their customers into different categories so as to ease the delivery of fixes as per the revenue they generate. This classification can be used as an additive factor to defect dependency metric as a process so that the quality teams can have this adopted and have it implemented to run regular business. Whenever two or more defects share a similar metric value or no value i.e. 0 or 1, these additive factors help product managers to take execution decision. It is quite normal to have defects with similar defect dependency value as they are backed on dependency relationship build using a defect model.

\section{Industrial Pilot Study}
This sections provides details on implementation of the above presented model based approach on few real-time pilot software projects used in day-to-day business by internal users of a software firm. The primary author of this paper is a consultant to the software firm and also pursing graduate studies on part-time basis. Due to non-disclosure clauses, the name of the firm is withheld. This study setup was conducted from December 2015 to March 2016. As part of our study, we have used \textit{i*} modeling framework to calculate defect dependency.

\subsection{ Understanding {\subsecit i*} modeling framework}

The \textit{i*} - Intentional Strategic Actor Relationships (iStar) modeling framework is a modeling technique developed especially to describe and analyze the dependencies among the entities of a software product during early and late requirement phase. It is a best fit modeling technique for products with continuous re-releases of products with incremental requirements during every maintenance and service pack release. In contrast to UML, \textit{i*} modeling framework provides a clear understanding on organizational relation in domain specific design. It was first formulated by E. Yu et al. \cite{Eric} as a modeling framework. The primary goal is to produce a simple model which is capable enough to help software practitioners to record relationships and dependencies among actors which is considered as a central concept of intentional actor. Fundamental concepts of \textit{i*} modeling framework are intended to represent requirements using desirable modeling characteristics, improve the readability, understand-ability, usability, feasibility of design models so as to enhance the overall consistency and effectiveness of the \textit{i*} modeling process. It also helps to understand the levels of complexity and provides scalability to an existing product to simply its work-flow within. Strategic Dependency (SD) model and Strategic Rationale (SR) model are two main modeling components in \textit{i*} modeling framework which are used to generate models to study the system design of a product. These models are usually designed using design guidelines specified for each and every entity to meet specific business requirement. These guidelines are unique to each model and have precise meaning \cite{Cares} behind every object which is diagrammatically suggested to practice.

\textit{Strategic Dependency (SD) Model} is a graph, which describes set of nodes and links where each node represents an actor and each link between two actors indicates that one actor depends on the other for something in order that the former may attain some goal. The SD model is used to express the network of intentional, strategic relationships among actors. SD diagrams depict the strategic dependencies between actors, but do not depict the internal rational behind these dependencies. Strategic Dependencies like Goal Dependency, Task Dependency, Resource Dependency, Soft goal Dependency and One-side dependency can be visualized using this model.

\textit{Strategic Rationale (SR) Model} is a graph, with several types of nodes and links that work together to provide a representational structure for expressing the rationales behind dependencies. It is considered to be an extension to SD model.

The actors with the SD model are ``opened up" to show their specific intentions. There are four types of nodes, based on the distinctions made for depended types in the SD model: goal, task, resource, and soft goal. There are three main classes of links internal to the \textit{i*} actor: means-ends links, task decomposition links and contribution links. SR diagrams open up actors and show all the internal elements that contribute towards analysing alternatives and fulfillment of the dependencies. Goals and Soft goals can be attributed to not only human actors, but also to non-human Actors (systems, machines, etc.) by the humans.

Both of these models are useful to generate goal dependencies and can also be used to generate use cases for testing. This helps us to calculate the network of dependencies for a defect in the \textit{i*} model. The modeling framework is now popularly adopted in cross domain research so as to meet functional and business requirements. E. Yu et al. proposed a relationship based applications for Process Re-Engineering of software products \cite{Yu}. Xavier et al. have proposed a framework for metric based actor-dependency model \cite{Xavier} for requirement engineering.  Gemma et al. proposed an architectural analysis tool \cite{REDEPEND} to study the dependencies at architectural level to understand current and future scalability options for existing software products. In this paper, we will be utilizing Strategical Dependency (SD) model to construct a quality model so as to study the defect widespread in a large scale software products. 

\subsection{Understanding istarxml file}
Once we obtain the defect flow of the reported defect, we will have to review the istarml file which is the XML representation of the strategic dependency (SD) model to understand the types of dependencies available in the model.  Gemma Grau et al. have clearly postulated the methodologies \cite{Franch} on building the \textit{i*} strategic dependency (SD) models. The SD model consists of set of nodes which represent actors involved and a set of dependencies that describe relation between these actors. The actor (depender) that depends on some other (dependum) in order to meet a defined objective. Here the dependum can be a resource, task, goal and a softgoal which have designated meaning specified as per \textit{i*} modeling standard. The SD model can be constructed using \textit{i*} tools like OpenOME, OME, REDEPEND-REACT-BCN \cite{REDEPEND} etc. which can be used either as a plugin or as a desktop based modeling tool. Below is a simple SR model for a Stock Business portal and its istarml representation. Here User and Stock Data are actors, View Stock Information is a resource available for the user to view. Hence User is a depender and Stock data is a dependum. Similarly Stock Data stored is a goal stored by user with Stock Data actor, where Stock data is a depender and User becomes a dependum.  
\begin{figure}[h]
\centering
\includegraphics [width=1\linewidth, height=4cm] {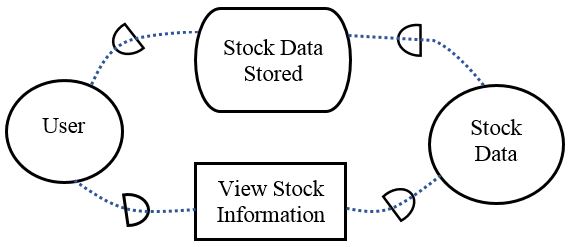}
\caption{User-Stock Data SD Model}
\label{sd}
\end{figure}

Below is the istarml XML file version of above defined simple SD model for user-stock application in Fig:\ref{sd}.
\begin{lstlisting}[style=XML]
<?xml version="1.0" encoding="UTF-8"?>
<istarml version=``1.0">
<diagram name=``Stock Info">
<ielement type=``goal" id=``_ZgNKVIe1xYa" name=``Stock Data Stored">
<ielement type=``resource" id=``_r2d2BB3TwArF" name=``View Stock Information">
<actor type=``role" id=``_T3outX21pQD" name=``User">
<dependency>
<depender iref=``_ZgNKVIe1xYa" aref=``_r2d2BB3TwArF">
<graphic content=``SVG"/>
</depender>
<dependee iref=``_ZgNKVIe1xYa" aref=``_r2d2BB3TwArF">
<graphic content=``SVG"/>
</dependee>
</actor>
<actor type="role" id="_LrmG117xey" name="Stock Data">
<dependency>
<depender iref=``_r2d2BB3TwArF" aref=``_ZgNKVIe1xYa">
<graphic content=``SVG"/>
</depender>
<dependee iref=``_r2d2BB3TwArF" aref=``_ZgNKVIe1xYa">
<graphic content=``SVG"/>
</dependee>
</actor>
</diagram>
</istarml>
\end{lstlisting}

Actors are represented using \textbf{$<$/actor$>$} tag. Dependency is enclosed between tag \textbf{$<$/dependency$>$}. Dependee are represented using \textbf{$<$/dependee$>$} tag and Dependum as \textbf{$<$/depender$>$}. Goal, Resource, Task and Softgoal are defined as \textbf{$<$/ielement$>$} with a unique identifier and name.

\subsection{Implementation}
This model based approach was executed on small pilot projects which were replacing the old tool used by internal engineers at a software firm. We first formulated the new requirement and build the scope of the implementation setup. This is to identify the challenges involved while setting up the ecosystem for this approach.

Table \ref{projs} contains the details of internal small scale pilot projects build for internal users along with their line of code. \textbf{\textit{RT}} is an customized request tracker tool designed as per internal requirements. \textbf{\textit{VMLab}} is an internal tool which is intented to manage and maintain Virtual Machines build for Testing and Development purposes. Both of these projects are re-constructed from existing legacy software and had to be rebuild to support latest business requirements. They were re-designed primarily due to their compatibility and performance issues. In regards to implementation, the focus on these two projects is to understand the behaviors of the approach on existing reconstruction projects. \textbf{\textit{MongoLD}} is a new tool designed as extension to the existing MongoDB database farm. The intent of this tool is to schedule jobs and perform regular daily data imports, exports and data extracts on previously constructed MongoDB database farm. The focus on applying this approach on this project is to understand the behavior of the approach on a new software project.

Most of the functional requirements of these projects were altered and modified during the re-design phase of development process. The defects recorded during development process were recorded and are tracked so as to estimate the cost and effort spent by developers and testers who were involved as part of these projects. The progress of these projects, development cycles, build cycle and defect fix cycle are managed and monitored by the lead developer of these projects independently. However, the defect dependency study and defect prioritization was performed by the Project Owners.
 
\begin{table}[h] 
\centering
\caption{Pilot Projects developed}
\label{projs}
\begin{center}
\begin{tabular} {|c|p{4cm}|c|}\hline
\textbf{Project Name}&\textbf{About Project}&\textbf{LOC}\\ \hline
\textit{RT}&Java based issue tracking tool used internally&12761\\ \hline
\textit{MongoLD}&Java tool to manage large data imports/exports into MongoDB farm&23679\\ \hline
\textit{VMlab}&Python tool to manage Virtual machines&12421\\ \hline
\end{tabular}
\end{center}
\end{table}

\subsection{Execution Work-flow}
\begin{figure}[h] 
\centering
\includegraphics [width=5cm, height=7cm] {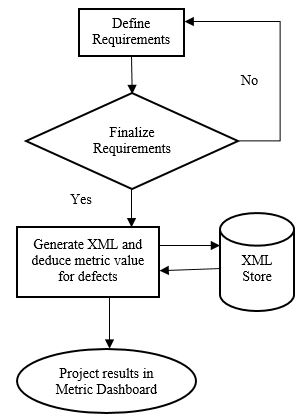}
\caption{Execution Work-flow of model approach}
\label{wrk}
\end{figure}

The common execution work-flow defined in Fig. \ref{wrk} for these small scale projects so as to decrease the cost, effort and maintenance challenges. The requirements are manually recorded by the Lead developer and are finalized by obtaining approvals from the business stake-owners. Once the requirements are finalized, lead developer would design the strategic dependency (SD) model for the entire product and its istarml XML files are stored in the MongoDB database (a document database). The unique ids of actors in istarml files are mapped similar to that of class names so that the istarml files of a defect can be programmatically extracted from the database. Implementation was setup using a Java code to calculate the metric information and is projected on a metric dashboard to understand the metric data. The developed projects are tested by a Quality Assurance (QA) teams and defects are logged with developers using an existing internal defect tracking tool. Lead developer will be using the defect dependency metric dashboard to prioritize the defects. This cycle is repeated until there is a halt to the incoming defects. Below Table \ref{deft} contains the number of defects reported and addressed until March 31 2016. These projects were successfully delivered and are operational to end users.

\begin{table}[h]
\centering
\caption{Total defects recorded and fixed}
\label{deft}
\begin{center}
\begin{tabular} {|c|c|c|}\hline
\textbf{Project Name}&\textbf{Total Defects}&\textbf{Defects Fixed}\\ \hline
\textit{RT}&519&499\\ \hline
\textit{MongoLD}&254&236\\ \hline
\textit{VMlab}&148&148\\ \hline
\end{tabular}
\end{center}
\end{table}

\subsection{Observations}
In this section, we captured interesting observations identified during the implementation of this approach on sample projects. These observations were specifically captured by collaborating with developers and testers involved in the execution of this model driven approach.
\subsubsection{Key Takeaways}
Below are the key takeaways recorded during the study:
\begin{itemize}
\item This approach is easy to adopt and simple to implement.
\item The implementation setup can be re-used for all new or existing projects.
\item Product stake-owners should be clear while defining the requirements as it has direct impact on calculating the metric value.
\item The calculated metric value might change w.r.t to the changes made to strategic dependency (SD) model of overall product. Hence it is recommended to finalize the requirements before the approach is implemented.
\item This metric can comprehend the health of a software product in terms of software quality.
\item Additional model analysis (parsing, transformation) is required to understand the actual dependencies across defect models.
\end{itemize}
\subsubsection{Challenges}
Below are the challenges recorded during implementation setup which might help large scale software products to ease their execution:
\begin{itemize}
\item There could be performance issues if the istarml XML files are physically read. It is advised to store the istarml XML files in a document databases like MongoDB/ Solr/ OrientDB etc. (open source) so as to improve the performance of metric calculation.
\item Configure a validator for istarml XML so as to check if the file is syntactically correct before applying the metric. This is to avoid incorrect results.
\item Frequent changes in requirements might alter the metric value for a specific defect. In such cases, the open defects are to be recalculated against latest SD model so as to obtain the latest actual value.
\end{itemize}

\section{Threats to Validity}
The Software Modeling framework is the basis of our proposed approach. We used  \textit{i*} (Intentional Strategic Actor Relationships) modeling framework to validate our process. We may not be able to calculate the defect dependency without creating a model XML for both recorded defect and for the overall product. It is required to have resources equipped with enough knowledge on tools used as part of building requirement models for their products. This is to create desired model XMLs in real time software ecosystems. As part of our pilot implementation, we were only able to test this approach on small software which are internally used within a software firm for smaller target audience ($<$300 users). These projects are not large enough to understand the real-time issues and possible improvements required to address the software quality in a larger software product. The end user base of small software products are not critical and sensitive enough when compared with larger products. Large Scale integrated software products have huge end-user base who come up with new business requirements after every version release. Product Managers are responsible for delivering those features without any data and functional loss. The real impact of our work can only be realized if our approach is applied at a larger scale either on an integration project or on a reconstruction project of any enterprise software producer. However, our implementation setup on these pilot projects helped us to gain clarity on standard functional setup and helped understand common challenges involved during our fresh start.

Dependency is a natural element among software entities. There is nothing wrong to have a real time code with high dependencies or high cohesiveness of code. In context to the recorded defect, large scale integrated software products suffer huge dependency issues than small software products. This is because of the complexity of their code flow and due to lack of clear understanding on control flow of the product. The most common process challenges recorded in real time integration and reconstruction projects are to migrate legacy products with the existing large scale software. Most of the time, product managers aren't sure on why a specific requirement was defined in a given legacy software. This creates a gap on having the current requirements mapped with alternatives in a new target software. This could be either due to lack of knowledge transfer or due to lack of requirement documentation. If the requirements and its dependencies are not formulated, our approach might not yield genuine results. The primary expectation of this approach is to understand the requirements with higher clarity so as to achieve fruitful results upon execution.

\section{Conclusion and Future work}
We consider model based defect dependency approach as an effective approach for new product manufacturers. Software developers can perform defect widespread analysis with a model as a strong basis. Over a period of time, the fully evolved product model will help quality engineers to estimate the defect widespread with higher accuracy and helps product stake-owners to create a stable product even after integrating new sub-products into the large scale integrated software product. As part of our validation, we used \textit{i*} modeling framework. However, we suggest Product Managers to adopt a new modeling language to construct a new work-flow defect models to improve integration bugs or use different modeling languages for different sub-products and later merge the defect dependency values for diverse results. 

As part of our future work, we will be working towards building a generic meta-model for defect widespread analysis in large scale software product development. We are also looking towards implementing model parsing and model transformation techniques to identify dependencies in defect model defined for a large software product and validate our approach on a enterprise large scale integrated software product for more authentic results.

\section{Acknowledgments}
We thank Service Delivery Team and Product Managers at FactSet Research Systems and SumTotal Inc. for supporting our research and sharing their experiences along with their valuable feedback.

\end{document}